\documentclass[english]{article}
\usepackage[T1]{fontenc}
\usepackage[utf8]{inputenc}
\usepackage[margin=1in]{geometry}

\usepackage{lmodern}
\usepackage[authoryear]{natbib}

\usepackage{setspace}
\usepackage{amsmath}%
\usepackage{amsfonts}%
\usepackage{amsthm}%
\usepackage[nolist]{acronym}
\begin{acronym}
  \acro{CI}{confidence interval}
  \acro{RCT}{randomized controlled trial}
  \acro{IV}{instrumental variable}
  \acro{LATE}{local average treatment effect}
  \acro{ATE}{average treatment effect}
  \acro{NBER}{National Bureau of Economic Research}
  \acro{OLS}{ordinary least squares}
  \acro{RD}{regression discontinuity}
  \acro{MTE}{marginal treatment effect}
\end{acronym}

\usepackage[bottom]{footmisc}
\usepackage{babel}
\newcommand\Perp{\protect\mathpalette{\protect\independenT}{\perp}}
\def\independenT#1#2{\mathrel{\rlap{$#1#2$}\mkern2mu{#1#2}}}

\title{Labour by Design:\texorpdfstring{\\}{} Contributions of David Card, Joshua Angrist, and Guido Imbens\thanks{Contact:
    \href{mailto:peter_hull@brown.edu}{peter\_hull@brown.edu},
    \href{mailto:mkolesar@princeton.edu}{mkolesar@princeton.edu}, and
     \href{mailto:crwalters@econ.berkeley.edu}{crwalters@econ.berkeley.edu}.
     We thank the editors of the \emph{Scandinavian Journal of Economics},
     Erik Lindqvist 
     and Andreas Moxnes, 
     for helpful feedback.
     We are grateful to Alberto Abadie, 
     Joshua Angrist, 
     Orley Ashenfelter, 
     David Card, 
     Guido Imbens,
     and Pat Kline 
     for excellent suggestions and comments.}}
\author{Peter Hull\\Brown University \and Michal Kolesár\\Princeton University \and Christopher Walters\\UC Berkeley}

\date{\today}

\usepackage{xcolor}%
\definecolor{webbrown}{rgb}{.6,0,0}
\usepackage{hyperref}%
\hypersetup{%
  breaklinks = true,%
  colorlinks = true,%
  anchorcolor = webbrown,%
  citecolor = webbrown,%
  filecolor = webbrown,%
  linkcolor = webbrown,%
  menucolor = webbrown,%
  urlcolor= webbrown,%
  citebordercolor= 1 0 0,%
  menubordercolor=1 0 0,%
  urlbordercolor=1 0 0,%
  runbordercolor=1 0 0,}
\usepackage{cleveref}

\begin{document}

\maketitle
\begin{abstract}
\noindent The 2021 Sveriges Riksbank Prize in Economic Sciences in Memory of Alfred Nobel was awarded to David Card ``for his empirical contributions to labour economics'' and to Joshua Angrist and Guido Imbens ``for their methodological contributions to the analysis of causal relationships.'' We survey these contributions of the three laureates, and discuss how their empirical and methodological insights transformed the modern practice of applied microeconomics. By emphasizing research design and formalizing the causal content of different econometric procedures, the laureates shed new light on key questions in labour economics and advanced a robust toolkit for empirical analyses across many fields.
\end{abstract}

\clearpage

\section{Introduction}

The 2021 Sveriges Riksbank Prize in Economic Sciences in Memory of Alfred Nobel was awarded to David Card ``for his empirical contributions to labour economics'' and to Joshua Angrist and Guido Imbens ``for their methodological contributions to the analysis of causal relationships.'' To an observer unfamiliar with modern microeconomic research, the connection between these two prize rationales may seem unclear. What does empirical labour economics, which studies the functioning of labour markets and institutions, have to do with methods for inferring causal relationships? Within economics, however, it's now taken for granted that robust empirical study---both inside and outside labour---often rests on a compelling approach for establishing causality. This widespread view reflects the profound impact of the three laureates on the field.

At its core, much of microeconomic theory concerns causal relationships. A good's own-price supply or demand elasticity represents the causal effect of a price increase on its quantity produced by firms or desired by consumers. Human capital theory \citep{becker_hc} explains the causal effect of schooling on workers' earning potential. The efficient design of labour market interventions, such as a new training program or a minimum wage increase, hinges on their causal effects on wages and employment. Yet for much of the 20th century, explicit causal language in microeconomic research was relatively rare. As late as the 1980s, only around 15\% of \ac{NBER} working papers contained the terms ``causality'' or ``causal'' \citep{ckz20}. But this share increased sharply in the mid-1990's and early 2000s, and has continued to grow. Now, nearly 50\% of \ac{NBER} papers feature causal inquiry \citep{imbens22}. What explains this change?

While it is challenging to convincingly answer such a (causal) question from a single time series, we can hypothesize one potential explanation from the timing of this change. Much of David Card's pathbreaking empirical work was conducted in the late 1980s and early 1990s, at the Industrial Relations Section of Princeton University. This work brought fresh causal evidence on core questions in labour economics: the employment effects of the minimum wage, the impacts of immigration on labour market outcomes of natives, and the effect of educational investments on labour market outcomes. The answers that Card's research uncovered were surprising and compelling, challenging conventional wisdom and triggering both debate and many follow-up analyses. Analyzing the impact of a large influx of Cuban migrants into Miami following the Mariel Boatlift, for example, \citet{mariel} found little effect on the employment rates and wages of native workers. Studying the effects of an increase in the New Jersey minimum wage, \citet{cardkrueger1994} found no evidence for a decline in low-wage employment. Both findings were at odds with simple textbook models of labour markets, and fuelled much subsequent empirical and theoretical research on potential explanations---such as the substitutability of foreign and native workers and the market power of low-wage firms. In this way, Card's work illustrated how surprising empirical facts---viewed as specific causal effects---could move strong theoretical prior beliefs and drive decades of new research.

But the impact of David Card's work on empirical practice stems not only from  what conclusions he reached but from \emph{how} he reached them. Card---along with Joshua Angrist, Orley Ashenfelter, Alan Krueger, other colleagues and students at the Princeton Industrial Relations Section, and other prominent labour economists at the time---pioneered a fresh approach to empirical analysis in the late 1980s and early 1990s: one centered on the identification of causal effects. The foundation of this approach is a careful consideration of a study's underlying \emph{research design}: an understanding of where the variation in an economic ``treatment,'' such as high minimum wages, came from, and an empirical approach that leverages this understanding to construct an appropriate comparison group. Desirable treatment variation often comes from so-called ``natural experiments''---unanticipated shocks or as-good-as-random shifts in the exposure to treatment, or in the factors determining treatment. This design-based approach can make transparent the key assumptions that drive an empirical study's conclusion, and often guides their empirical validation or falsification.

The design-based approach is compelling in part because of the close connections it draws to true experimentation, such as in a \ac{RCT}.\footnote{Card attributes his use of the term ``research design'' to his exposure to the \emph{New England Journal of Medicine}, which Alan Krueger subscribed to at Princeton and which often used the term in reference to randomized trials \citep{ck_interview}.} True randomization is often infeasible for studying important economic questions, such as the effect of immigration or effects of large-scale minimum wage changes on local labour markets. Most economic treatments are not just determined by chance, as in an \ac{RCT}, but also by individual or institutional choices which are far from random. Earlier solutions to the threat of \emph{selection bias} in such settings focused on models of  choice, drawing on a rich body of microeconomic theory. By instead focusing on the quasi-randomness in certain natural experiments, Card and fellow pioneers of the design-based approach showed how such modeling restrictions could be relaxed or even eschewed with by-chance variation.

The design-based approach was made more convincing and rigorous by new econometric insights, recognized in the second half of the 2021 Nobel Prize. In the celebrated \ac{LATE} theorem of \cite{ia94}, the laureates showed how a natural (or true) experiment generating randomness in a variable influencing treatment could be leveraged to estimate the causal effects of the treatment---with minimal restrictions on other factors influencing the treatment choice. For example, a randomly drawn lottery number that determines the eligibility of an individual for military draft service may be used to estimate the effects of such service on later-life earnings \citep{angrist90}. But some individuals may volunteer in the military regardless of draft eligibility, while others may find ways of avoiding military service when drafted. Economists typically analyze such settings with \ac{IV} techniques, using draft eligibility as an ``instrument'' for the military service ``treatment.'' Imbens and Angrist showed that such \ac{IV} analysis generally recover a \ac{LATE}: the average earnings effect of military service among \emph{compliers}, who are induced to service as a result of the draft lottery. In contrast, a hypothetical \ac{RCT} that randomly assigns people to serve in the military would recover the overall average treatment effect in the entire population (including volunteers and those who avoid the draft regardless of their lottery number), not just among compliers.

This central insight of the \ac{LATE} theorem had a profound effect on how economists interpret evidence produced by natural experiments, and how the field synthesizes evidence accumulated across different studies. Underlying the theorem is a potential outcomes framework, which relaxed the model-based (and often parametric) restrictions from earlier analyses of IV. The framework highlighted the potential for treatment effect \emph{heterogeneity}, both across populations and across different natural experiments in the same population. Such heterogeneity explains how the results from one research design may differ from another, despite both yielding valid causal effects. The focus on treatment effect heterogeneity and the associated notions of internal validity vs. external validity (i.e. generalizability) has motivated a vast and growing literature---including later work by the laureates---showing how structural models of individual behavior and statistical extrapolations can synthesize causal evidence across different research designs and settings.

This article argues that the face of modern empirical economics was shaped in large part by the empirical and methodological contributions of the 2021 Nobel laureates. Card showed how careful attention to research design can bring new, compelling, and sometimes unexpected evidence to core questions and theories in labour economics. Angrist and Imbens showed the precise strengths and limitations of the design-based approach, allowing new empirical work to be better contextualized and integrated across studies. Together, the laureates strengthened the scientific foundation of economics by showing how robust theory and empirics can interact to advance our understanding of key economic questions where true experimentation is infeasible.

To put the contributions of Card, Angrist, and Imbens in context, we first outline the key empirical challenge the laureates' work focused on---selection bias---and the state of contemporary empirical practice when this work began.  We discuss how the design-based approach and search for natural experiments helped to address several critiques of existing empirical approaches. We then turn to the empirical contributions of the laureates, focusing on David Card's work on immigration and minimum wage laws as well as the three laureates' work on the earnings effects of education and labour market experiences. Next, we detail the methodological piece of the prize, centered around the Angrist and Imbens LATE theorem. We discuss how their result and the general potential outcome framework formalized key strengths and limitations of the design-based approach for IV and led to further insights for other methods. We conclude by briefly summarizing other advances, by the laureates and others, that grew out of this prize-winning work.

\section{Setting the Stage}\label{sec:background}

\subsection{The Selection Challenge}
Many important questions in economics hinge on the reliable measurement of causal effects. To decide whether to expand a subsidized post-schooling training program, a policymaker must weigh the effects of the program on trainee labour market outcomes against the costs of the program and the likely effects of alternative programs, such as job-search assistance.  When considering healthcare reform, a social planner must take into account the effects of health insurance on individual health and welfare as well as the social cost. Should the national minimum wage be raised to \$15 an hour? The answer depends, in part, on the likely causal effect of such an increase on low-wage unemployment.

Unfortunately, the answers to such causal questions tend to not be directly revealed by economic data. A researcher cannot simply compare individuals who are ``treated'' (those who participate in a training program, who have health insurance, or who are subject to a higher local minimum wage law) with individuals who are ``untreated'' to estimate the treatment's effects, since these two populations are likely very different. Individuals can choose whether to participate in a training program, and this choice may be driven by a wide range of characteristics and circumstances that are relevant for the observed outcomes. Trainees tend to have lower levels of education and past earnings than non-trainees, for example, and differences in both schooling and work history may signal underlying human capital differences. Thus, even if the causal effect of a training program is positive, a researcher may find that trainee earnings after completing the program are lower than non-trainee earnings.

This problem of \emph{selection bias} has long been recognized in economics. A conceptually simple solution is to remove the element of treatment choice with a randomized experiment.  This ``gold standard'' for evaluating medical treatments (and some social programs) purges bias by ensuring the individuals randomized into the treatment and control groups are, on average, identical prior to the experiment. Unfortunately, such randomization may be difficult or infeasible for many important economic questions. It is difficult to imagine researchers convincing government officials to randomly raise the minimum wage in some areas but not others or to experiment with other high-stakes economic programs.

In response to this fundamental identification challenge, economists have developed a variety of econometric methods that aim to purge selection bias in non-experimental settings by incorporating additional data and assumptions. The simplest of these approaches is to adjust for observable differences between the groups, often with linear regression. Another approach is to leverage longitudinal data, for example by comparing the earnings of individuals before and after their participation in a training program. By further contrasting the treatment group's earning change with an analogous change in the untreated group, one arrives at what \citet{AsCa85} termed a ``difference-in-differences'' analysis. In both cases, regression-adjusting or time-differencing may address selection bias by making the treated and untreated group more comparable. A different and clever strategy, pioneered by James Heckman (\citeyear{heckman74,heckman76,heckman79}), leverages microeconomic theory to model an individual's decision to self-select into treatment and derives a statistical bias correction from the selection equation. This model-based approach grew out of a long tradition of simultaneous equation modeling in economics and quickly caught on in the early 1980s (see \citet{blundell01} for a review).\footnote{Heckman was awarded the Nobel Memorial Prize in 2000, alongside Daniel McFadden, for ``his development of theory and methods for analyzing selective samples.''}

\subsection{Empirical Concerns}

Despite decades of empirical work leveraging regression adjustment, longitudinal data, and selection models to answer important questions in labour economics, by the mid-1980's there was growing concern that the collective evidence produced in this work was weak. In a review of studies of the union wage gap, for example, \citet{lewis86handbook,lewis86book} documented a range of estimates so large as to be of little use. He further showed that estimates constructed from elaborate selection models appeared even less reliable than simpler regression estimates, noting that  ``a substantial fraction of [selection method] estimates are\ldots preposterously large or outlandishly negative'' \citep[p.~1144]{lewis86handbook}. In a staff study of training programs for the congressional joint economic committee, \citet[p.~14]{goldstein72} called for improving the evaluation process for assessing training programs, since ``the robust expenditures (\$179.4 million from fiscal 1962 through 1972)\ldots are a disturbing contrast to the anemic set of conclusive and reliable findings.'' This assessment is echoed in \citet{ashenfelter78}, which documented a possible reason for the unreliability: trainees' earnings tend to fall prior to joining the program, in both absolute terms and relative to a comparison group of non-trainees. This form of self-selection into treatment, dubbed ``Ashenfelter's dip,'' suggests that part of the observed earnings increase following training may reflect mean-reversion---i.e., a return to a permanent earnings path that was temporarily disrupted. Simple longitudinal methods are likely to ascribe any such increase to the effect of the treatment, making the program appear more effective than it actually was.

Broadly, these critiques suggested that empirical studies using observational data and existing econometric methods rarely solve the selection challenge in labour economics, and may thus not be a satisfactory substitute for randomized experiments. A direct assessment of the extent to which such studies can reproduce experimental evidence was given in a landmark study by \citet{lalonde86}. LaLonde first calculated the effect on trainee earnings of the National Supported Work Demonstration, an employment program that randomized participation in a field experiment. He then compared the estimates to those produced by a variety of non-experimental methods based on a modified data set, where the experimental control group  was replaced by different comparison groups drawn from the Panel Study of Income Dynamics (PSID) and a Matched Current Population Survey---Social Security Administration (CPS---SSA) File. In spite of using state-of-the-art econometric methods, LaLonde failed to replicate the experimental results without the experimental control group. Different methods and comparison groups produced a wide range of estimates, and standard specification tests were unhelpful in determining which observational estimates were closest to the experimental ``ground truth.''

These findings suggested that more and better data, though clearly necessary, were not sufficient for credible causal inference. This conclusion exposed cracks in the foundation of an argument made in \citet{stafford86}, who documented a stark rise in the share of empirical labour economics papers in 1965-1983 using individual-level data (like the PSID and CPS) instead of aggregate data on census tracts, states, or countries. Stafford argued that the granularity of such ``microdata'' protected labour economics from the critiques that had been leveled at empirical economics as a whole, by \citet{leamer83} and others.\footnote{Other prominent critiques are found in \citet{hendry80}, \citet{sims80}, \citet{black82}, and \citet{leontief82}.} But the core of Leamer's critique was that many empirical findings were sensitive to small changes in the analytic assumptions---exactly the concern raised by the above studies. Granularity of microdata alone, as it turns out, doesn't make a study robust. 

Leamer worried that such a lack of robustness encouraged specification searches, in which researchers tinkered with the analytic method they used until they found a desired result. The proposed remedy was sensitivity analysis, in which researchers show how their results change with the exact specification or functional form (or, in a Bayesian analysis, by varying the prior distribution). Such robustness checks are now widely used in economic studies. But while sensitivity analyses can reveal the limitations of an observational analysis, they rarely suggest solutions on their own. Many carefully executed papers at the time, such as the longitudinal studies of training programs by \citet{ashenfelter78} and \citet{AsCa85}, were already upfront about the fragility of their results.

LaLonde's analysis also suggested one path forward: putting less emphasis on model-based solutions to self-selection into treatment, which did not clearly dominate simpler regression-based approaches in his study, and more emphasis on the researcher's choice of the comparison group. While ``the difficulty of obtaining a reliable comparison group'' \citep{ashenfelter74} had long been noted in labour economics, LaLonde's analysis made clear that sensitivity to this choice could be as or more important than the particular econometric approach. Furthermore, LaLonde showed that it can be hard to determine from the data alone which comparison group or method is most likely to address selection bias. Addressing the prevailing concerns in applied microeconomic research would seem to require help from elsewhere.

\subsection{Labour Economics by Design}

A primary contribution of the 2021 laureates was to push the field towards approaching causal questions in a fundamentally different way: with an emphasis on \emph{research design} as a means to address the sensitivity concerns raised by \cite{ashenfelter78}, \cite{lewis86handbook,lewis86book}, \cite{lalonde86}, and others. While exact definitions of research design vary (and sometimes overlap with other terms, like ``empirical strategy'' or ``identification strategy''), it broadly refers to a researcher's understanding of the process determining how units in a study are assigned to different ``treatments,'' or the process determining their outcomes in the absence of a treatment---which can, in turn, be used to construct a sensible non-experimental ``control group''  \citep{meyer95}. From this perspective, a convincing study makes core bias concerns explicit through a clear discussion of the research design, which in turn dictates the appropriate econometric methods. The researcher provides direct or indirect evidence supporting the key assumptions underlying the methods---often loosely called the ``identifying assumptions.'' The laureates' empirical work exemplifies this push towards research design in the early 1990s, as we discuss in \Cref{sec:princeton}. Their later methodological work formalized the design-based approach, as we discuss in \Cref{sec:ImbensAngrist}.

An emphasis on research design brings new perspective to the argument of \citet{stafford86}, on the virtues of microdata in empirical labour economics. First, while more detailed data may allow researchers to probe new sources of variation and consider different identifying assumptions, granular data does not by itself make for a convincing study. To leverage and validate a particular research design, one needs data on the appropriate variables---which need not be contained in standard datasets. Sometimes such data must be collected by the researcher, as in the seminal \citet{cardkrueger1994} study discussed below. In other cases a design calls for linking different administrative data to more standard research extracts, as in the \citet{angrist90} study also discussed below. New data and linkages could, in turn, enable new lines of research. The rise of such researcher- and administratively-collected data in applied microeconomics, over the more standard microdata research extracts emphasized in \citet{stafford86}, clearly coincides with the increase in design-based research \citep{angristpischke2009,ckz20}.

A design-based perspective also brings insights to \citeauthor{leamer83}'s (\citeyear{leamer83}) sensitivity critique. Identifying assumptions, which involve restrictions on the treatment assignment process or comparability of outcomes for treatment and control groups, are emphasized and distinguished from other choices in the estimation procedure. The ``parallel trends'' assumption underlying the \citet{cardkrueger1994} difference-in-differences analysis, or the key \ac{IV} assumptions of random assignment, exclusion, monotonicity, and relevance in the \citet{ia94} framework, are presented and scrutinized. Other parts of estimation, like exactly how the researcher controls for covariates or weights different subpopulations, are less central when they are not informed by the research design. The effect of such choices can be assessed by more routine sensitivity analysis, while the plausibility of identifying assumptions may require institutional knowledge to assess. This hierarchy of assumptions reflects an emphasis on design, with transparent treatment-control comparisons over potentially complex models and statistical procedures. A state-of-the-art non-parametric IV estimator based on an instrument with unclear assignment may be less convincing than a simple comparison of trends in a well-executed difference-in-differences analysis. As \citet{rubin08} puts it: ``[for] causal inference, design trumps analysis.''

The design-based approach advanced by the laureates also raises new issues for empirical practice. While a clear and plausible research design may lead to \emph{internally} valid estimates of causal effects, which are free from selection bias in a given study population, the \emph{external} validity (i.e. generalizability) of such estimates to other populations and contexts is far from guaranteed. Similarly, the search for treatments with clear assignment processes or plausible control groups may limit the scope of microeconomic research. Some questions are more easily cast in a design-based approach, while others may seem fully out of reach by concerning treatments that are not easily viewed as manipulable by any design. We return to these and other issues in \Cref{sec:ImbensAngrist}.

\subsection{Natural Experiments}

Where do convincing research designs come from? The most obvious source is a \ac{RCT}, with true experimental assignment. \acp{RCT} may be a natural way to answer some questions in economics, like the effectiveness of job training programs (as argued in \citet{AsCa85} and \citet{ashenfelter87}). Indeed, the past 30 years have seen a steep rise of field experiments through microeconomics \citep{levitt/list:09}. Prominent examples range from randomly offering housing vouchers to allow low-income households to move to better neighborhoods through the Moving to Opportunity Program \citep{KlLiKa07,ChHeKa16};to randomly offering Medicaid insurance coverage to low-income adults in the Oregon Health Insurance Experiment \citep{oregon1,oregon2}; to entrepreneurial researchers setting up their own charity to randomize how a charity solicits donations in a study of the determinants of charitable giving \citep{dlm12}. The field of development economics, in particular, has seen a stunning transformation in the share of experimental papers, which include evaluating the effect of educational policies such as deworming children \citep{MiKr04} or changing teaching incentives \citep{gik10}, studying the effects of microcredit provision \citep{bdgk15}, or the rates of return to fertilizer \citep{dkr08,dkr11}.\footnote{Three of the scholars leading this transformation---Abhijit Banerjee, Esther Duflo, and Michael Kremer---were awarded the Nobel Memorial Prize in 2019, ``for their experimental approach to alleviating global poverty.''} In a randomized controlled trial, the research design is clear and often under the researcher's control. Validating the identifying assumption of random assignment is also straightforward: if units are assigned to different treatment randomly, observable pre-treatment characteristics should be unrelated to treatment status. Such balance can be checked by testing if the distribution of covariates is the same across the treatment arms.

Much of the work by the 2021 laureates, however, considers economic questions where direct experimentation is rare or infeasible. Here a convincing research design may come from a natural or quasi-experiment, in which some institutional quirk or force of nature generates variation that is plausibly as-good-as-randomly assigned or which otherwise suggests appropriate treatment and control groups.\footnote{\cite{Dinardo2020} differentiates natural experiments as ``serendipitous randomized trails,'' where a variable of interest is as-good-as-randomly assigned across units, and quasi-experiments which rely on parallel trends assumptions or other restrictions on the comparability of unobservables across treatment and control groups. However, like ``research design,'' exact definitions of these terms vary---see, e.g., \cite{Titiunik2020} for alternative definitions and discussion.}  Such variation may take the form of large but unforeseen shocks to regions or markets. \cite{freeman89} was an early proponent of basing empirical analyses around such large shocks, illustrating the value of this approach in an analysis of the federal minimum wage imposition on Puerto Rico in the 1970s \citep{CFFr92}, and in an analysis of a post-Sputnik boom in the demand for physicists in the U.S. \citep{freeman75}. Other quasi-experimental analyses leverage more narrow variation, often across individuals in the same region or market, where the research design arises from idiosyncrasies in the rules used to administer some economic variable. An example of this approach is the regression discontinuity design of \citet{ThCa60}, where a threshold assignment rule (such as a minimum test score for students to avoid taking remedial classes) can be leveraged to study the effect of assignment among individuals just above and just below the threshold. \citet{campbell69} was an early proponent of using such discontinuities and other quasi-experimental designs in psychology.\footnote{The review by \citet{meyer95} helped to make economists aware of this quasi-experimental tradition in psychology.} Variation in the timing of policy shocks across regions, such as U.S. states, may also provide a basis for a compelling research design: \citet{gruber94} exploited such variation in an influential study of the incidence of mandated maternity benefits.

Labour economists were early adopters of natural experiments as the foundation of a research design. This work included studies by \citet{RoWo80}, \citet{mvd95}, and a group of researchers at the Industrial Relations Section (part of the Princeton University Economics department). Gary Solon's work \citep{solon85} showed how one can leverage changes in laws as a source of variation. Orley Ashenfelter and Alan Krueger famously combined a random quirk of nature with innovative data collection to study the returns to education in \citet{AsKr94}. The researchers  traveled to the annual Twins Days Festival (in Twinsburg, Ohio) to gather data on pairs of identical twins. Independently asking both twins about each other's schooling levels allowed them to account for measurement error in self-reported years of schooling---a prominent concern in existing returns to schooling studies. By looking at how differences in twins' education levels predict differences in twins' earnings, Ashenfelter and Krueger were further able to address concerns of ability bias in existing estimates. Here the identifying assumption follows from the natural experiment of twinning, which generates two individuals with identical genetics and thus (arguably) comparable earnings potential. Leveraging the more narrow variation in education within twin pairs, Ashenfelter and Krueger found much greater labour market returns than in previous cross-sectional studies which adjusted for observable demographics and family characteristics.

Two other prominent members of the Industrial Relations Section are two of the 2021 laureates: David Card and Joshua Angrist. We next discuss how their work made both empirical contributions to key questions in labour economics and methodological advances in using design-based approaches to answer them.

\section{Empirical Landmarks}\label{sec:princeton}

Empirical work by the laureates is distinguished by its clarity and compelling analysis of several important topics. We organize a partial review of this work into four broad topics: the effects of immigration on local labour markets, the effects of minimum wages on low-wage employment, the labour market returns to schooling, and other determinants of labour earnings.

\subsection{Effects of Immigration on Local Labour Markets}

Immigration raises several important and hotly debated questions in economics and public policy. A key policy concern in many countries is that relaxed immigration laws can induce large inflows of migrants, sometimes with low levels of education and labour market experience, to compete with local workers and potentially reduce wages and employment prospects. Theoretical predictions about the effects of such  low-skilled labour supply shocks to local labour markets are ambiguous, since they depend on many factors---including the substitutability of immigrant and native labour, existing labour market institutions, and the potential response in labour and product demand. Early empirical studies of this question typically used cross-sectional variation in immigrant populations to estimate production function parameters that speak to these mechanisms \citep[e.g.,][]{grossman82}. As noted by \cite{borjas87}, however, the self-selection of immigrants to different labour markets may bias such estimates.

Two early studies by Card (\citealp{mariel}, and \citealp{altonjicard91}) show how an increased focus on research design and large-scale natural experiments can address such selection concerns. In \citet{mariel}, the so-called 1980 Mariel boatlift was used to study the local labour market effects of low-skill immigration in Miami. This large and arguably unanticipated labour market shock grew out of rising political unrest in Cuba, and led to an unprecedented influx of immigrants in the Miami labour market. The boatlift was announced on April 20th, 1980, with Fidel Castro allowing anyone wishing to emigrate from Cuba to leave through the port of Mariel. Between April and October 1980, nearly 125,000 people left Cuba via a flotilla of private vessels. Roughly half settled in Miami, where several processing camps had been established.

Three aspects of the Mariel episode made for a compelling natural experiment. First, the source of the immigration shock was clear and external: the political turmoil which gave rise to the boatlift was plausibly unrelated to other factors affecting labour markets in the United States. Second, the scale of the shock was enormous: the boatlift led to a 7\% increase in the size of Miami's total labour force over the span of a few months. The Mariel immigrants tended to be less educated and worked in lower-skill occupations than the overall Miami population, implying an even larger increase in labour supply at the lower end of the skill distribution. Third, large survey data sets contained the variables needed to study this large and external shock: baseline labour market conditions immediately prior to the boatlift could be measured in the 1980 census, while the CPS included reasonably large samples for measuring effects in subsequent years. Importantly, unlike most other ethnic groups, Cubans are separately identified in the CPS---allowing Card to study outcomes for both Cubans and other Hispanic immigrants.

Despite these advantages, the appropriate way to leverage the Mariel boatlift experiment to study immigration effects is not immediately clear. Miami is only one city, and neither the timing nor the location of the immigration shock were truly random. Moreover, the onset of the 1982 recession in the aftermath of the Mariel boatlift highlighted the possibility that the effects of the immigration shock could be confounded by changes in other national or regional labour market trends.

To allay this concern, Card constructed a comparison group that might plausibly account for these other time-varying factors and isolate the impact of immigration. Specifically, Card compared Miami's labour market trajectory to those of Atlanta, Los Angeles, Houston, and Tampa-St. Petersburg---a group of cities that featured roughly similar demographics and exhibited similar trends to Miami prior to the boatlift. The idea of forming a control group to adjust for time-varying confounders in a non-experimental setting grew out of Card's earlier work on longitudinal earnings models and training programs \citep{AsCa85,abowdcard1985}.

The \cite{mariel} analysis revealed that wages and unemployment moved similarly in Miami and the comparison cities from before to after the boatlift, including for lower-skilled groups of workers. This suggests that the large influx of Mariel immigrants had limited impacts on native outcomes, a surprising finding that spawned a large literature on the channels through which local labour markets adjust to immigration. Methodologically, the \cite{mariel} study provided a clear example of how to combine a natural experiment with a carefully-constructed control group to produce compelling empirical findings.\footnote{The synthetic control method \citep{AbGa03,adh10synthetic}, which combines multiple control groups to construct a single synthetic control mimicking the treatment group, can be seen as a further refinement of this idea.} This paper presaged the growth of difference-in-differences studies, which have since become one of the most common empirical strategies in applied microeconomics.\footnote{\cite{ckz20} find that nearly a quarter of \ac{NBER} working papers in 2020 employed difference-in-differences, constituting around 60\% of all \ac{NBER} working papers using an experimental or quasi-experimental approach.}

The findings in \cite{mariel} were striking, and not without critique.\footnote{For recent discussion, see \cite{borjas2017} and \cite{periyasenov2018}.} One clear concern was generalizability: the Mariel experiment was a large shock, but its effects were concentrated in one arguably unique labour market. Indeed, \cite{mariel} notes that Miami's long history of receiving Cuban immigrants may complicate the interpretation of the findings. For a more comprehensive view of local labour market effects, \cite{altonjicard91} famously devised an \ac{IV} strategy which translated the logic of the \cite{mariel} research design to a national level. Just as Cuban immigrants tended to locate to cities like Miami, where there were large groups of previous Cuban immigrants, immigrants from other countries tend to settle in regions with existing immigrant enclaves. Altonji and Card used previous immigrant settlement patterns to instrument for immigration to different metropolitan areas, finding large negative effects of immigrant inflows on native wages but no effect on employment. Later, \cite{card2001} refined this \ac{IV} strategy with a ``shift-share'' instrument that predicted inflows by city and occupational groups. This shift-share approach is now widely used to study the effects of immigration and other treatments combining large external shocks (e.g. immigrant inflows) and heterogeneous local exposure (e.g. immigrant enclaves).\footnote{Shift-share, or ``Bartik'' instruments can be traced back to \cite{freeman80}, \cite{Bartik1991}, and \cite{Blanchard1992}. A recent methodological literature, including \cite{gss20}, \cite{bhj22}, and \cite{akm19}, formalizes how such instruments can leverage quasi-experimental variation.}

The design-based approach to immigration study has gained immense popularity in the years following \cite{mariel} and \cite{altonjicard91}, and has generated several strands of literature seeking to explain why immigration has limited effects on the labour market outcomes of native workers in some---but not all settings (see \cite{dss16} for a recent review). Key sources of heterogeneity appear to include the distribution of native skill (particularly communication skills; e.g. \citealp{ps09}) and the ease of technological adjustment (e.g. \citealp{dg15}). While there is still ongoing debate over the magnitude of wage and employment effects from immigration overall, such studies of heterogeneity and mechanisms are no doubt helped by a clearer understanding of research design.

\subsection{Effects of Minimum Wages on Low-Wage Employment}

A similarly important and fiercely debated question in economics and public policy is the effects of federal or local wage floors on low-wage employment. The textbook model of a perfectly competitive labour market predicts that an increase in the minimum wage results in movement along a downward-sloping market demand curve for labour, creating unemployment and reducing worked hours among the employed. By the early 1990s, the conventional view among labour economists was that these predictions accurately describe the causal impacts of changes in minimum wage laws. It had long been recognized that increasing the minimum wage can theoretically boost employment by flattening the supply curve facing an employer with market power \citep{robinsonbook}, but this scenario was generally regarded as specific to situations with a single monopsonistic employer and irrelevant to the functioning of typical low-wage labour markets. Empirical evidence from the 1970s and 1980s, largely based on time-series or cross-sectional variation in minimum wages across states, was broadly consistent with the competitive view (see \citealp{brown1982review} and \citealp{cardkrueger1995review}
for reviews).

\citet{cardkrueger1994} famously revisited the effects of the minimum wage using a natural experiment derived from an increase in New Jersey's state minimum wage. Their strategy built on earlier work by \citet{card1992ca}---who analyzed the effects of a minimum wage increase in California using a comparison set of unaffected states---as well as \cite{katzkrueger1992}, who studied an increase in the federal minimum wage by comparing establishments paying higher vs. lower wages prior to the change (see also \citealp{card92fed}). In 1990, the New Jersey legislature passed a law that would increase the state's minimum wage from \$4.25 to \$5.05 per hour as of April 1, 1992. Anticipating this change, \cite{cardkrueger1994} surveyed a set of fast food establishments on both sides of the New Jersey/Pennsylvania border immediately prior to the change (February-March 1992) and again a few months afterward (November-December 1992). Combining elements of the \cite{card1992ca} and \cite{katzkrueger1992} strategies, \cite{cardkrueger1994} compared changes in outcomes in New Jersey to those in Pennsylvania, as well as changes in outcomes for restaurants with higher vs. lower baseline wages within New Jersey.

Like the Mariel boatlift analysis, the New Jersey/Pennsylvania study exhibits several hallmarks of modern studies using natural experiments in applied microeconomics. The action in the variable being studied (the minimum wage) originated from a specific and interpretable source (the New Jersey law change) rather than uncontrolled state-level variation of unclear origin. The size of the shock was large: New Jersey's minimum wage increased by roughly 20\%, and most fast food restaurants in New Jersey paid below \$5.05 before the change. Careful attention was paid to constructing and validating a control group that could plausibly capture the counterfactual path of outcomes in the absence of treatment for affected units. The controls here consisted both of restaurants across the border in Pennsylvania and of high-wage New Jersey restaurants less exposed to the reform. Finally, \cite{cardkrueger1994} assembled detailed microdata to measure outcomes for the treatment and control groups. In this case, rather than relying on existing surveys, they fielded their own custom survey instrument tailored to the question at hand.

Card and Krueger's baseline survey in February/March 1992 showed roughly similar wage distributions in New Jersey and Pennsylvania, with average starting wages just over \$4.60 and about one-third of restaurants in each state paying exactly the baseline minimum wage of \$4.25. By the endline survey in late 1992, about 90\% of New Jersey restaurants reported paying exactly the new minimum of \$5.05 while the wage distribution in Pennsylvania appeared roughly unchanged. But despite this large differential change in wages, Card and Krueger's survey showed no evidence of a negative employment impact on New Jersey restaurants. In fact, average full time employees (FTE) at New Jersey stores increased slightly, while average FTEs in Pennsylvania fell, resulting in a modest positive difference-in-differences estimate. An ``exposure design'' comparing high- and low-wage employers within New Jersey likewise showed a small relative increase in employment at low wage restaurants. In contrast to the textbook perfectly competitive model and the earlier time-series evidence, the Card and Krueger empirical results suggested that increasing the minimum wage did not reduce employment---and if anything may have increased it.

The empirical findings of \cite{cardkrueger1994} upended conventional wisdom on the effects of minimum wages, generating backlash in some quarters of labour economics. Despite the well-known theoretical result that minimum wages could increase employment in settings with employer market power, adherents of the competitive view of labour markets derided the \cite{cardkrueger1994} findings as unscientific (see, e.g., \citealt{buchananwsj}). While negative, such reactions highlight the value of natural experiments and careful research design. In contrast to empirical work from earlier eras, in which \cite{leamer83} argued that ``hardly anyone takes anyone else's data analysis seriously,'' the results from a compelling natural experiment call out for explanation and further study even among skeptics of the substantive conclusions. As it turns out, the conclusions of several recent studies are broadly consistent with the \cite{cardkrueger1994} finding of limited effects of the minimum wage on employment (\citealp{dube2010}, \citealp{cengiz2019}, \citealp{giuliano2013}, \citealp{hl2019}, and \citealp{dustmannqje2022}). Partially motivated by these findings, an increasing body of work has investigated competitive structure and monopsony power in labour markets (see \citealp{cchk2018} and \citealp{manning2021}, for two recent reviews). Renewed interest in employer monopsony power---following \cite{cardkrueger1994}, the extended treatment in \cite{ck_book}, and the analysis of \citet{manning_book}---has since fueled a large literature on firm wage-setting (see, e.g., recent work by \citealp{ams20,klms20,lms20} and \citealp{bhm22}).

\subsection{Effects of Schooling and Experience on Earnings}

A long literature in economics and related fields considers the effects of education and labour market experience on subsequent earnings and employment. While the theoretical impact of increased human capital is unambiguous (e.g. \citealp{becker_hc}), the empirical evidence for such causal effects was limited throughout most of the 20th century. The \cite{coleman66} report famously showed in cross-sectional regressions that the fraction of variance in student achievement attributable to educational inputs was small relative to the contribution of family background. Surveying the large empirical literature following the Coleman report, \cite{hanushek86} concluded that there was virtually no relationship between educational inputs and subsequent outcomes. Of course, selection bias looms large for such studies as the deployment of educational resources to students and schools is far from random.

Two influential studies by David Card and Alan Krueger (\citealp{ck1992quality,ck1992gaps}) investigated the effects of school quality on labour market outcomes by isolating a clever source of variation: the movement of students across different U.S. regions. \cite{ck1992quality} estimated returns to schooling separately by cohort and state of birth, controlling for cohort-specific state of birth and state of residence effects in the 1980 US census. This strategy compares relationships between earnings and schooling for individuals in the same birth cohort working in the same state but educated in different states, leveraging cross-state moves to measure differences in cohort-specific school quality across states. Card and Krueger related these estimated returns to measures of school quality for each state and birth cohort, showing that school quality improvements such as reduced pupil/teacher ratios appear to increase the return to education. To study the role of school quality in the evolution of the Black-white wage gap, \cite{ck1992gaps} estimated separate returns to schooling by race, state of birth, state of residence, and birth cohort. Between cohorts born in the 1920s and the 1940s, they documented a striking relative increase in the return to education for Southern-born Black men compared both to non-Southern-born Black men and to Southern-born white men within regions of residence. The timing of this differential change in returns coincided with a relative increase in measures of school quality for Southern-born Black men, suggesting an important role for school quality in reducing the racial wage gap over time. Though these studies did not take advantage of a sharp policy change, they effectively used individuals moving between locations as a collection of natural experiments---removing permanent differences between locations of birth to flexibly account for unobservables. Recent work in several areas builds on this idea of using ``movers'' to mitigate selection bias, including studies of firm effects \citep{akm1999,chk2013,cck2016}, neighborhood quality \citep{chettyhendren2018part1}, and variation in regional healthcare utilization \citep{finkelstein2017movers}.\footnote{Earlier examples of papers using such designs to study industry wage differentials include \cite{murphy1987unemployment}, \cite{krueger1988efficiency}, and \cite{gibbons1992does}.}

The Card and Krueger studies suggest a non-zero ``return to schooling:'' the theoretical parameter governing causal effects of increased education on labour market earnings. Perhaps the most famous estimates of the returns to schooling from this period comes from \cite{angristkrueger1991}, which used a creative \ac{IV} strategy---based on an institutional quirk of the U.S. education system---to address the clear self-selection issue of earlier regression-based analyses.\footnote{Another seminal contribution is \cite{card95}, who used the distance to nearby colleges in an individual's birthplace as an instrument for educational attainment.} Children in the U.S. traditionally start first grade in the calendar year in which they turn six, but most state compulsory schooling laws allowed students to drop out on their sixteenth birthday. The combination of these two rules implies that compulsory schooling laws are more stringent for students born early in the calendar year. Consider, for example, two children born in 1930 with one born in January and the other born in December. These children would likely be in the same schooling cohort, starting first grade together in the fall of 1936. The individual born in January would be among the oldest of his or her classmates, reaching the compulsory school age of 16 in January 1946, in the middle of 10th grade. The child born in December would be among the youngest in the class and be compelled to stay in school until the middle of 11th grade. If both students drop out as soon as the law allows, the December child will attain nearly a full year of additional schooling than the child born in January.

This argument suggests a novel instrument for years of schooling: a child's birthday. The interaction between age-at-entry and compulsory schooling rules suggests that birthdays may affect educational attainment. Moreover, it seems plausible that birthdays are as good as randomly assigned, and have no effects on earnings through channels other than completed schooling.\footnote{\cite{buckleshungerman2013} note that maternal characteristics vary with birthday in recent birth cohorts, suggesting that birthdays may not be fully independent of family background.} \cite{angristkrueger1991} operationalized this idea using instruments based on season (quarter) of birth, the measure of birthday available in public-use decennial census data.\footnote{With more detailed data on date of birth, this strategy can be sharpened into a regression discontinuity design leveraging the shift in school entry dates for children born immediately before and after the turn of the calendar year. An example of this approach appears in \cite{clarkroyer2013}.} 

Angrist and Krueger's analysis revealed that the relationship between birthday and education is evident in the data for men born in the 1920s through the 1940s. On average, children born in the second through fourth quarters of the year stay in school one tenth of a year longer than those born in the first quarter. Angrist and Krueger presented a battery of falsification exercises suggesting that this pattern is due to their proposed compulsory schooling mechanism. For example, using point-in-time school enrollment for teenagers in the 1960 and 1970 censuses, they demonstrated that the gap in enrollment between first- and later-quarter births emerges at age 16 only in states where the compulsory school-leaving age is 16 rather than 17 or 18. This can be seen as an early example of the placebo and robustness checks that are now commonly used to probe identifying assumptions in design-based studies.\footnote{This exercise strengthens the case for a causal interpretation of the quarter-of-birth first stage. In traditional simultaneous equations models, a causal interpretation of the first stage is unnecessary -- the first stage is a linear projection and any bias stems from a relationship between the instrument and unobservables in the outcome equation. In the design-based view, however, it is seen as unlikely that an instrument is as-good-as-randomly assigned unless both the first stage and reduced form are free of selection bias. This idea is made explicit in the framework of \cite{ia94}, which features a causal model of the first stage as detailed in Section \ref{sec:ImbensAngrist}.} In addition to attaining 0.1 fewer years of schooling, individuals born in the first quarter of the year also earn about one percent less than those born later. \ac{IV} estimates formed as the ratio of these two differences (as discussed more in Section \ref{sec:late}) imply that a year of schooling boosts earnings by roughly 10 percent.

Methodologically, the \cite{angristkrueger1991} analysis differed from several of the above design-based studies of immigration and the minimum wage in two important ways. First, while the  \cite{mariel} and \cite{cardkrueger1994} studies looked at large and sudden shocks to aggregate labour markets, \cite{angristkrueger1991} leveraged narrow birthdate variation across individuals within markets. Second, while the 1980 Mariel boatlift and 1992 New Jersey minimum wage change were paired with natural comparison groups, unlike with birthdates it is difficult to imagine these deliberate policy changes as occurring by chance. The narrow and plausibly as-good-as-randomly assigned \ac{IV} variation in \cite{angristkrueger1991} thus marked a key methodological shift in the use of natural experiments in economics while, as we discuss below, highlighting new econometric questions.

Substantively, the studies by Card, Angrist, and Krueger helped change the consensus on the effects of educational investment, from the rather pessimistic conclusion of the \cite{coleman66} report to the modern consensus that school resources generally matter (see \citealp{jackson2020} for a recent review). As with the immigration and minimum wage literatures, recent work focuses on the heterogeneity of educational input effects across settings and individuals. \ac{IV}-based analyses of charter school effects and variation in school quality within urban districts (\citealp{kipp_aer}, \citealp{aadkp:11}, \citealp{aahp:16}, \citealp{vam}) is one area where Angrist has continued to study such heterogeneity.\footnote{Other notable examples include \cite{angristlavy99}, which uses regression discontinuity to study class size effects, and the \ac{RCT} of \cite{angristetal02} which studied private school vouchers.}

\subsection{Other Determinants of Earnings}

Three other studies by the laureates (\citealp{angrist90}, \citealp{imbens2001}, and \citealp{CaHy05}) are worth highlighting, for the creative use of naturally occurring randomization to answer important questions on the determinants of labour market earnings. In \cite{angrist90}, the random assignment of draft lottery numbers was used to study the effects of Vietnam-era military service on earnings. Paralleling the issues in the \cite{lalonde86} analysis of training program effects, military veterans differ from non-veterans on many dimensions, and earlier efforts to address this selection with the available econometric tools yielded unstable and inconclusive estimates. \cite{angrist90} leveraged public lotteries that assigned Random Sequences Numbers (RSNs) to dates of birth for men born between 1950 and 1955. For the 1950-1952 birth cohorts, men whose RSNs fell below a cutoff were conscripted into military service.\footnote{As \cite{angrist90} notes, RSNs also generated an increase in military service for those born in 1953 even though this cohort was never drafted, as men with low lottery numbers preemptively enlisted to improve their terms of service in anticipation of the possibility of conscription.} Randomly-assigned draft lottery numbers are clearly independent of earnings potential and plausibly affect outcomes only through military service, making the draft an attractive natural experiment for studying service effects on labour market outcomes.\footnote{Earlier work by \cite{hearst1986} used the Vietnam-era draft lottery to study effects of draft eligibility on mortality.}

\cite{angrist90} obtained a custom version of the Social Security Administration's Continuous Work History Sample (CWHS) augmented with birthdates in order to link earnings to draft RSNs. Veteran status was only partially determined by RSNs, however: some men enlisted regardless of lottery number, while many with low lottery numbers did not serve due to deferrals or performance on pre-induction mental and physical screening tests. As a result, the difference in military service rates for eligible and ineligible men was only around 15 percentage points. This non-compliance calls for an \ac{IV}, with the RSN serving as instrument for veteran status. \cite{angrist90} divided the difference in mean earnings by draft eligibility in the CWHS by the difference in service rates in the Survey of Income and Program Participation (SIPP) to construct IV estimates of the causal impact of military service. The results showed that military service led to a 15 percent earnings penalty for the Vietnam draft cohorts.

The \cite{imbens2001} study likewise uses randomization afforded by a lottery to shed light on a different question: what are the wage effects of non-labour income? The effect of unearned income on economic behavior is a foundational question in labour and public economics but is difficult to measure since non-labour income is likely correlated with many unobserved determinants of labour supply and other outcomes. To address this, \cite{imbens2001} conducted a special survey of Massachusetts lottery players. The sampling frame for the survey took advantage of the fact that the state maintains historical records of lottery winners, including some individuals that won millions of dollars and some that won small amounts. Winners of small prizes provide a natural control group for bigger winners. To lend support to the key identifying assumption that the magnitude of the prize is as good as randomly assigned, \cite{imbens2001} conducted balance checks which showed no correlation between the prize magnitude and individual characteristics (e.g. prior earnings) once the winners of the largest prizes are excluded. Their analysis revealed modest negative effects of unearned income on labour earnings, with somewhat larger impacts for older workers. Recent studies of the impacts of unearned income have followed the \cite{imbens2001} lottery-based approach \citep[see, e.g.,][]{cesarini2017}.

Like these two studies, \cite{CaHy05} leveraged natural randomization to study the wage effects of a large-scale program: the Self Sufficiency Project (SSP), a Canadian program which made earnings subsidies available for a random pool of long-term welfare recipients over three years. Such effects speak to a large literature in labour and public economics considering possible employment disincentives from mean-tested welfare programs, such as the Earned Income Tax Credit (EITC). Unlike the EITC and earnings subsidies in other countries, the SSP was only available for full-time work; participants furthermore had to establish eligibility by working full-time within the first year of program participation. \cite{CaHy05} showed how this program structure created distinct incentives to find a full-time job in the first year and to continue working once eligbility was established. To tease apart these channels, they developed and estimated a dynamic model with the experimental variation in program participation. Their estimates showed that the combination of ``establishment'' and ``entitlement'' incentives generated a striking pattern in the experimental effects, which peaked in the second year following random assignment before fading. Notably, there were no long-run effects on either wages or welfare participation, suggesting temporary wage subsidies may not induce program dependency. Beyond these substantive findings, the \cite{CaHy05} approach to estimate structural models by exploiting natural experiments helped push the frontier of the design-based approach, as we discuss more below.

\subsection{Taking Stock}

These and other early studies of natural experiments produced new and compelling evidence on several classic questions in labour economics. At the same time, they often raised new questions about how such evidence is best interpreted and synthesized into a broader body of scientific knowledge. The increased emphasis on the forces determining the assignment of certain economic ``treatments'' highlighted that design-based studies often leverage highly specific sources of variation. What does the \citet{cardkrueger1994} result for fast-food workers in New Jersey teach us about the impact of minimum wage more broadly? Is the lack of labour market effects from a large immigration shock in \cite{mariel} specific to the 1980s Miami labour market? These questions of interpretation loom especially large in the \ac{IV} analysis of \cite{angristkrueger1991}, where the identifying variation in individual birthdates led to relatively small differences in completed education: comparing the schooling of individuals born in the first and fourth quarter of a year suggests that at most 10\% were on the margin of dropping out as soon as they are legally allowed. To what extent can this narrow source of variation inform the returns to schooling in the general population?

The interpretation of design-based \ac{IV} estimates of the returns to schooling was carefully considered in an influential review by \citet{card99}. He noted that such estimates---including in \cite{angristkrueger1991}---typically exceed corresponding \ac{OLS} estimates, often by 30\% or more. This pattern would seem to present a puzzle, since standard selection bias reasoning suggests that \ac{OLS} estimates should be biased \emph{upward}, not downward, as students with higher unobserved ability are likely to select more schooling. While measurement error in self-reported years of education could explain some of the discrepancy, as it would tend to attenuate the \ac{OLS} estimates, it is unlikely to explain the often large gap between \ac{IV} and \ac{OLS} estimates. To close this gap, \citet{card99} offered another explanation: the sub-populations shifted into treatment by the variation in the quarter of birth or other instrumental variable strategies may have higher returns to schooling than the overall population. This idea of heterogeneous causal effects driving the interpretation of \ac{IV} estimates was formalized in the seminal analysis of \cite{ia94}.

\section{A Deeper Understanding of Causality}\label{sec:ImbensAngrist}

Interpreting estimates of conceptually similar causal effects across different natural experiments and designs requires a flexible econometric framework. Specifically, it requires a way to think about how the specific source of identifying variation might affect the interpretation of the quantity being estimated. Consider, for example, the Vietnam draft study of \citet{angrist90}, where draft eligibility (i.e. an RSN below the conscription cutoff) was used to instrument for military service. Most individuals who served in Vietnam were volunteers who would have served no matter their RSN number. Presumably, the draft-based identification strategy of \cite{angrist90} cannot speak to the effect of serving in the military for these volunteers. But in what formal sense is this true?

A standard way of motivating the use of \ac{IV} methods in such applications is the potential for selection (or ``omitted variables'') bias: individuals who do and do not serve in the military differ in many observed and unobserved ways, some of which may affect their adult earnings. One way to address this concern is to model the selection process:  \citet{heckman74,heckman76,heckman79}, \citet{heckmanrobb85}, \cite{chamberlain86}, and others showed how \ac{IV} could identify such selection models through a combination of exclusion and functional form restrictions.\footnote{Other work showed how average effects can be bounded without such restrictions: see, e.g., \cite{robins89}, \cite{manski90}, and \cite{balke_pearl}} Selection models can also be used to structure heterogeneity across different instruments and samples---at least when its clear what observable and unobservable characteristics are relevant and in what way. But this may be hard to do in a flexible manner. More importantly, a model-based approach to bias and heterogeneity may obscure the advantage of a randomly assigned instrument as in \cite{angrist90}.

A key innovation in \citet{ia94} is to approach the selection bias problem and \ac{IV} solution from a different direction. Instead of deriving model restrictions sufficient to fully correct selection, and align the \ac{IV} analysis with a hypothetical randomized experiment, Imbens and Angrist asked what minimal assumptions make a simple linear \ac{IV} estimand causally interpretable. Their answer helped separate the conceptual roles of ``chance'' (i.e. as-good-as-random instrument assignment) and ``choice'' (assumptions on the selection process) in design-based analyses, clarified how different quasi-experimental studies of conceptually similar economic quantities could be synthesized, and highlighted more general strengths and weaknesses of the design-based approach.

\subsection{The Potential Outcomes Framework and LATEs}\label{sec:late}

To keep the individual heterogeneity in the response to treatment unrestricted, \citet{ia94} cast the IV estimation problem in a potential outcomes framework. This framework dates back to \cite{neyman23}---who first proposed a version of it for analyzing randomized experiments---and \citet{rubin1974,rubin1978,rubin1990} who later generalized it for observational studies. The core logic of the potential outcome framework is also found in the early econometric literature, including work on IV as a method to solve simultaneous causality: \cite{wright28}, \cite{working27}, \cite{tinbergen30}, and \cite{haavelmo43} all distinguished between potential economic variables determined by structural relationships and the observed variables determined in market equilibria.\footnote{This early literature also can be seen as laying the foundation of later graphical formalizations of causality and related methods, particularly the path analysis method of \cite{wright28}. The do-calculus of Pearl (\citeyear{pearl95,pearl00,tbow18}) has evolved in parallel with the potential outcomes framework in recent years, though the latter generally remains more popular in applied economics. See \cite{heckmanpinto2014}, \cite{pearl15}, and \cite{imbens20} for recent discussions.} The distinction between the observed outcomes of simultaneous equations models (such as equilibrium prices and quantities) and the ``potential'' outcomes which might be realized under certain counterfactuals is especially clear in  \cite{haavelmo43}, who focused on the challenge of interpreting observed data on income and consumption in terms of parameters governing marginal propensities to consume and invest.\footnote{Trygve Haavelmo received the Nobel Memorial Prize in 1989, ``for his clarification of the probability theory foundations of econometrics and his analyses of simultaneous economic structures.'' Jan Tinbergen was awarded the first Nobel Memorial Prize in 1969, along with Ragnar Frisch, ``for having developed and applied dynamic models for the analysis of economic processes.''} The emphasis on potential outcomes was revived in the early 1990s by  \cite{heckman90}, \cite{manski90}, and others, along with \citet{ia94}; these papers showed the value of the clarity that explicit potential outcome notation delivers.\footnote{As \cite{imbens14} notes, much of the post-war econometric literature used a notation only involving realized or observed outcomes; see also \cite{hendrymorgan92} and \cite{imbens97} discussions of this history.}

To sketch the potential outcomes framework, consider the causal effect of some binary treatment $D_i$ (say, enlisting in the army) on some subsequent outcome $Y_i$ (say, adult earnings). We imagine two potential outcomes associated with the treatment, $Y_i(1)$ and $Y_i(0)$, representing the earnings of individual $i$ if they did and did not to enlist in the army. Only one of these potential outcomes is observed, depending on the value of $D_i$; the other is the individual's \emph{counterfactual} outcome, associated with the unrealized treatment state. Formally, the observed outcome can be written
\begin{equation}\label{eq:POs}
Y_i=(1-D_i)Y_i(0)+D_iY_i(1)=Y_i(0)+D_i\left(Y_i(1)-Y_i(0)\right),
\end{equation}
where the quantity $Y_i(1)-Y_i(0)$ represents the effect of $D_i$ on $Y_i$ for individual $i$.

When $D_i$ is randomly assigned, i.e., we were to randomly enlist some individuals in the military but not others, it becomes independent of the potential outcomes $Y_i(1)$ and $Y_i(0)$. This ensures that the \ac{ATE} is identified by the difference in average outcomes among treated and untreated individuals:
\begin{equation*}
    E[Y_i\mid D_i=1]-E[Y_i\mid D_i=0]=E[Y_i(1)\mid D_i=1]-E[Y_i(0)\mid D_i=0]=E[Y_i(1)-Y_i(0)],
\end{equation*}
where we use \Cref{eq:POs} in the first equality and the random assignment assumption in the second equality. The potential outcome notation makes the magic of a randomized experiment transparent: because we never observe both potential outcomes for each individual, we can't ever learn their individual treatment effect, $Y_i(1)-Y_i(0)$. But by virtue of random assignment, we can still learn the value of the treatment effect on average, in the population of interest.

To adapt the potential outcome framework to an IV setting, \citet{ia94} defined two sets of potential outcomes, one set for the treatment $D_i$ and one set for the outcome $Y_i$.\footnote{Footnote 2 in \cite{ia94} attributes the adoption of potential outcome notation for $D_i$ to Gary Chamberlain, who---along with Donald Rubin---were faculty at Harvard when Imbens and Angrist started there as assistant professors.} Specifically, let $D_i(z)$ be the potential treatment status when $Z_i=z$. In the \citet{angrist90} example, for instance, $D_i(0)$ is the military status of the individual if they were draft-ineligible, while $D_i(1)$ is the potential military status if they were draft eligible. Since both the treatment $D_i$ and the instrument $Z_i$ are manipulable, Imbens and Angrist defined potential outcomes $Y_i(d, z)$ over potential values $d$ of the treatment $D_i$, and potential values $z$ of the instrument $Z_i$. These correspond to earnings under each combination of serving in the military and draft eligibility.\footnote{\cite{ia94} derived the LATE theorem with a multivalued $Z_i$, but we focus on the case with binary $Z_i$ here for simplicity.} They then considered four substantive assumptions:
\begin{itemize}
    \item \emph{Random assignment:} $(Y_i(0,0), Y_i(1,1), Y_i(1,0), Y_i(0,1), D_i(0), D_i(1))\Perp Z_i$;
    \item \emph{Exclusion:} $\Pr\left(Y_i(d, 0)=Y_i(d, 1)\right)=1$ for each $d\in\{0,1\}$;
    \item \emph{Monotonicity:} $\Pr\left(D_i(1)\ge D_i(0)\right)=1$; and
    \item \emph{Relevance:} $\Pr\left(D_i(1)> D_i(0)\right)>0$.
\end{itemize}
The first assumption requires the instrument to be as-good-as-randomly assigned with respect to the potential outcomes and potential treatment choices. This assumption holds automatically for instruments such as the draft-eligibility instrument, or arguably the quarter-of-birth instrument in \citet{angristkrueger1991}. Under random assignment, we can estimate the average effect of the instrument on the treatment, $E[D_i(1)-D_i(0)]$, following the logic of randomized experiments above. By the same logic, we can also estimate the \emph{intent-to-treat effect}, the average effect on the outcome of switching the instrument from zero to one: the effect of being draft eligible on earnings.

The second assumption, or ``exclusion restriction,'' requires any effects of the instrument $Z_i$ on the outcome $Y_i$ to arise from changes in the treatment $D_i$: varying the instrument while holding the actual treatment fixed has no effect on the outcome. This condition allows us to define potential outcomes $Y_i(d)$ indexed by treatment status $d$ alone, like in the case of a randomly assigned treatment. The first two assumptions together capture the sense in which the instrument is ``exogenous''  in conventional IV analysis. The potential outcomes framework makes it clear that there are actually two separate assumptions underlying this condition. Even if the instrument is randomly assigned, it may fail the exclusion restriction if, for instance, draft-eligible individuals temporarily leave the country in order to avoid the draft and this dodge has an effect on later life earnings.

The third monotonicity assumption, most original to \citet{ia94}, requires the instrument to only affect treatment status in one direction: without loss of generality, we assume either that switching from $Z_i=0$ to $Z_i=1$ increases $D_i$ (i.e. $D_i(1)>D_i(0)$) or has no effect. In other words, being draft eligible weakly encourages everyone to serve in the military: no individuals would serve in the military if they were draft-ineligible, but refuse to serve if they were draft-eligible---a mild assumption in this case. To help interpret this assumption, \citet{air96} define four types of individuals, indexed by their potential treatments. First,  there are always-takers: individuals who volunteer to serve, regardless of their eligibility status, $D_i(1)=D_i(0)=1$.  Second, there are never-takers who avoid the draft,  $D_i(1)=D_i(0)=0$. Third, there are compliers: individuals who  serve only if they are draft-eligible, $D_i(1)=1$, $D_i(0)=0$. Finally, there could be defiers, who only serve if they are draft-\emph{ineligible}. The monotonicity assumption can be seen to rule out the presence of such unusual behavior. 

The final condition is a relevance assumption, which says that the fraction of compliers in the population is positive. In other words, there are people whose treatment status can be manipulated by changing the instrument. It is not difficult to come up with an instrument that satisfies the first three assumptions---flipping a coin for each individual would do---but finding an instrument that jointly satisfies all four assumption requires some ingenuity. Statistically, the relevance assumption ensures that the correlation between the treatment and the instrument is nonzero, $Cov(Z_i, D_i)\neq 0$. This ensures that we do not divide by zero in the definition of the IV estimand,  $\beta^{IV}=\frac{Cov(Z_i, Y_i)}{Cov(Z_i, D_i)}$.\footnote{Here $\beta^{IV}$ is the population analog of the \cite{wald40} estimator for a bivariate regression with mismeasured regressors (see \cite{angristpischke2009} for a discussion).}

Under these assumptions, \cite{ia94} showed that the IV estimand $\beta^{IV}$ identifies a local average treatment effect (LATE):
\begin{align*}
    \beta^{IV}=\frac{E[Y_i\mid Z_i=1]-E[Y_i\mid Z_i=0]}{E[D_i\mid Z_i=1]-E[D_i\mid Z_i=0]}=E[Y_i(1)-Y_i(0)\mid D_i(1)>D_i(0)],
\end{align*}
where the first equality follows from the definition of $\beta_{IV}$ and
the fact that $Z_i$ is binary.\footnote{For the second equality, note that by random assignment $E[D_i\mid Z_i=1]-E[D_i\mid Z_i=0]=E[D_i(1)-D_i(0)]$ and $E[Y_i\mid Z_i=1]-E[Y_i\mid Z_i=0]=E[Y(D_i(1))-Y_i(D_i(0))]$, following similar steps as in the above ATE identification proof. By monotonicity, $E[D_i(1)-D_i(0)]=\Pr(D_i(1)>D_i(0))$ and $E[Y(D_i(1))-Y_i(D_i(0))]=E[(Y_i(1)-Y_i(0))(D_i(1)-D_i(0))]=E[Y_i(1)-Y_i(0)\mid D_i(1)>D_i(0)]\times \Pr(D_i(1)>D_i(0))$, completing the proof.} The LATE $E[Y_i(1)-Y_i(0)\mid D_i(1)>D_i(0)]$ is a ``local'' treatment effect, since it corresponds to the average treatment effect for compliers only; in the \cite{angrist90} context, this is the average effect of military service on adult earnings among individuals whose draft lottery numbers compelled them to serve.

The Imbens and Angrist analysis delivers three key insights. First, it clarifies precisely how the source of variation in the treatment induced by the instrument affects the interpretation of the estimand $\beta^{IV}$. Second, it helps separate statistical assumptions (random assignment) from substantive economic restrictions (exclusion and monotonicity). Finally, it emphasized how causal interpretation of IV requires the treatment and the instrument need to be ``manipulable.'' We discuss each insight in turn.

\paragraph{Internal and External Validity}

The LATE result showed precisely how the quasi-experimental variation in an instrument affects the IV estimand when no structural restrictions are placed on the treatment effects. The estimand identifies an average effect for the compliers. If the IV leverages a ``narrow'' source of variation---as in \cite{angrist90} and \citet{angristkrueger1991}---then the group of compliers may be a small subset of the overall population. In \citet{angristkrueger1991}, the compliers are those who are on the margin of dropping out of school, but are induced to stay on for an additional year due to their exact quarter of birth. These individuals comprise at most 10\% of the overall population. Intuitively, we cannot ever learn from data alone about the treatment effect for never-takers (or always-takers), since we never see them treated (or untreated) in the data. While the identity of compliers is never directly given by data (since we never observe both $D_i(1)$ and $D_i(0)$ for the same individual $i$), \cite{abadiekappa} showed how a wide range of functions of their predetermined characteristics and potential outcomes could be estimated.

Relative to an experimental ideal, where we could learn the average treatment effect for the overall population, the LATE result may appear underwhelming. But the group of compliers is often of policy interest. For example, the \citet{angristkrueger1991} compliers may help inform policies that affect the minimum school-leaving age. It is true that the LATE is less relevant for predicting effects of other policies, such as the effect of abolishing college tuition. After all, people attending college---or those considering attendance---are not those who are affected by minimum schooling laws that the variation in quarter of birth leverages. Here the value of the LATE result lies in knowing that for considering the effects of such a policy, we need to combine the \citet{angristkrueger1991} estimates with an economic model that would allow us to extrapolate the treatment effect estimates to this population, or else look for a more informative natural experiment.

More generally, the LATE result sharpens the distinction between internal validity of the study (when can we interpret the IV estimate as the average treatment effect for compliers?) and its external validity (what are the lessons that carry over to other settings?).\footnote{\cite{campbell57} gives an early formalization of the difference between internal and external validity in the social sciences.} A concern raised in \citet{HeUr10} and \citet{deaton10} is that the increased use of natural experiments puts too much emphasis on internal over external validity and that too many studies stop at reporting the IV estimate, which may not answer a question of economic interest. As argued in \citet{imbens10}, the value of the LATE framework lies in separating the assumptions needed to identify the treatment effect for compliers in the current population from any additional assumptions needed to generalize the internally valid estimate to other populations. This allows for more transparency when researchers complement the quasi-experimental variation in the data with a structural model. For example, it allowed \citet{CaHy05} to combine experimental variation in an earnings subsidy with a structural model to identify the impact of the subsidy on welfare entry and exit rates. The \citet{imbens2001} study, also discussed in previous section, likewise combine experimental variation with a life-cycle model of labour supply. Recent work by \cite{brinch_et_al}, \cite{mst18}, and \cite{klinewalters2019} clarifies connections between the LATE framework and model-based identifying restrictions---and how the former can be used to relax the latter.

\paragraph{Statistical vs. Substantive Restrictions}

The second insight of the LATE result lies in separating the substantive restrictions in an IV analysis that always need to be justified by economic reasoning---the exclusion restriction and the monotonicity assumption---from the random assignment assumption which can hold automatically if the variation in the instrument is as good as random. This clarifies what exactly randomization delivers: it allows us to identify the intent-to-treat effect. But additional assumptions are needed to go from this effect to the treatment effect for compliers.

The distinction between the exclusion restriction and the random assignment assumption allows for a more nuanced analysis of potential violations of instrument ``exogeneity.'' Statistical balance checks can be used to verify that the randomization ``worked.'' In contrast, while statistical tests of the exclusion restriction and monotonicity exist \citep[see, e.g.,][]{kitagawa15}, most credible applications of IV rely on institutional or theoretical arguments to justify them. Such arguments are
often used to develop indirect application-specific diagnostic checks for these assumptions. This targeted probing of the design validity would not be possible if Imbens and Angrist did not make the assumptions clear in the first place.

Making the key assumptions clear also allows for more targeted sensitivity analysis. For example, \citet{air96} show that under violations of the monotonicity condition, $\beta^{IV}$ averages the treatment effect for compliers with the treatment effect for defiers, but the weight on the defiers is negative. On the one hand, this implies that $\beta^{IV}$ could be negative even if treatment effects are positive for all individuals. On the other hand, the result also implies that the presence of defiers is of lesser concern if their proportion is small, since the weight placed on them is proportional to the size of the defier group.\footnote{Similarly, certain violations of the exclusion restrictions may have little effect on the interpretation of the results, as explored, for example, in \citet{kcfgi15}. See also \cite{imbensrubin97} for a discussion of sensitivity analyses when the exclusion restriction and monotonicity assumptions are violated in a Bayesian framework.}

Another way of interpreting the \citet{ia94} assumptions is to think of them as an exploration of model misspecification. Early formalizations of instrumental variables methods focused on linear structural models for the outcome and supplemented it with an ``exogeneity'' assumption that the residual in this equation is uncorrelated with the instrument. The LATE result shows what happens when we drop the parametric restrictions.

\paragraph{The Role of Manipulation}

The potential outcomes framework generally highlights the need for manipulation in causal analyses: to interpret the potential outcomes $Y_i(d, z)$ and potential treatments $D_i(z)$, one needs to be able to manipulate, at least in principle and at least for some subpopulation, the treatment and the instrument. If the treatment is an innate attribute of a unit that cannot be manipulated, this notation makes it clear that we cannot speak of causal effects---in line with \citeauthor{rubin75}'s \citeyear{rubin75} dictum \citep[echoed in][]{holland86}: ``no causation without manipulation.''\footnote{As the recent literature on ``shift-share'' and related instruments show, different views on which components of a treatment or instrument are manipulable can lead to vastly different identifying assumptions, estimation concerns, and inferential procedures \citep{gss20,bhj22,akm19,borusyakhull}.} For instance, as discussed by \citet{GrRu11}, the notation makes it clear that while it is difficult to talk about the causal effect of sex or race, we can talk about causal effects of being perceived as having a certain sex or race. Such perception effects have been studied in evaluating the effects of blind auditions \citep{GoRo00}, or in countless ``audit studies'' that manipulate otherwise identical résumés---by, say, changing the name on the résumé (see, e.g., \cite{BeMu04} for an early example).

\subsection{Extensions and Connections}

While our exposition focused on the simplest setting with a binary treatment and a binary instrument, the framework extends readily to cases with multi-valued or multi-dimensional instruments (such as indicators for quarter of birth). \citet{AnIm95} consider settings with multi-valued treatment (such as years of education), demonstrating that in this case IV recovers a generalization of LATE known as the Average Causal Response (ACR). \citet{AbAnIm02} generalize the setup to cover estimation of quantile treatment effects.

In another important contribution, \citet{AnGrIm00} adapted the LATE framework to cover estimation of demand or supply elasticities in a simultaneous equation system of supply and demand. This is a classic problem that originally motivated the use of instruments by Philip and Sewell Wright, Tinbergen, Haavelmo and other early pioneers of IV methods. This work was extended substantially by the Cowles Commission, whose worked showed how exclusion and covariance restrictions can be used to identify two-equation supply and demand models, as well as more complicated simultaneous equations systems \citep{christ94}. To explain the identification challenge, suppose both the log of the demand curve $Q_i^d(P)$ and the log of the supply curve $Q_i^s(P)$ in market $i$ are linear in log of the price $P$:
\begin{align}
    \ln Q_i^d(P)&=\alpha^d+\beta^d \ln P+\varepsilon_i^d\label{eq:linear_demand},\\
    \ln Q_i^s(P)&=\alpha^s+\beta^s\ln  P+\varepsilon_i^s\label{eq:linear_supply},
\end{align}
where $\beta^d<0$ and $\beta^s>0$ are demand and supply elasticities, respectively (we assume these are constant across markets $i$), and $(\varepsilon^d_i, \varepsilon^s_i)$ are unobserved demand and supply shocks. Since equilibrium price equates supply and demand, both the observed equilibrium price $P_i$ and the equilibrium quantity $Q_i$ depend on the supply and demand shocks.\footnote{Specifically, setting supply equal to demand and solving for price yields $\ln Q_i=(\beta^d\alpha^s-\beta^s\alpha^d)/(\beta^d-\beta^s) + \frac{\beta^s}{\beta^s-\beta^d}\varepsilon_i^d -\frac{\beta^d}{\beta^s-\beta^d}\varepsilon_i^s$ and
$\ln P_i=(\alpha^s-\alpha^d)/(\beta^d-\beta^s)+\frac{1}{\beta^s-\beta^d}\varepsilon_i^d-\frac{1}{\beta^s-\beta^d}\varepsilon_i^s$.} As a result, a simple regression of observed log quantity on observed log price will recover neither the demand nor the supply elasticity, but a hard-to-interpret mixture of the two. The IV solution to this simultaneity challenge, as first considered by the Wrights, Tinbergen, and others, is to measure some component of the supply shock that does not affect demand. Due to this ``exclusion restriction,'' one can show that using such a supply shock component as an instrument for log price in a regression of log quantity on log price regression recovers the demand elasticity.\footnote{The use of the ``exclusion'' term in this context can be traced back at least as far as \cite{koopmans49}.} If we instead use a component of the demand shock that doesn't affect supply, we recover the supply elasticity.

But what if we relax the assumption that elasticities are constant across markets, and that the unobserved shocks are additive? \citet{AnGrIm00} consider a non-parametric setup that doesn't impose any functional form restrictions on the supply and demand curves $Q_i^d(P)$ and $Q_i^s(P)$. In this unrestricted setup, they show that an IV regression using an instrument that shifts the supply curve, but does not affect demand, identifies a weighted average of market-specific demand elasticities. If the demand elasticity varies with price, the estimand also averages over different prices in the same market. Like the LATE result in the context of estimating causal effects of a binary treatment, this result clarifies the role of internal and external validity of the IV estimates, and the role of functional form restrictions imposed in the classic linear model \eqref{eq:linear_demand}--\eqref{eq:linear_supply}.

The LATE framework has also been central to understanding \ac{RD} designs, a quasi-experimental design where treatment eligibility is determined by whether a particular variable, called a running variable, crosses a threshold. For example, to estimate the effects of class size on student test scores, \citet{angristlavy99} exploit the fact that class sizes in Israel follow the rule of Maimonides, a twelfth century rabbinic scholar: a school should not have class sizes bigger than 40. Here the running variable is the class size, and 40 represents the threshold. If a student cohort in a particular school comprises fewer than 40 students, they will all be in one large classroom. But if there are 41 students, they become eligible for a small classroom treatment: the school is allowed to open two classes with an average size of 20.5. If schools follow the Maimonides' rule exactly, we can estimate the effect of the small classroom treatment on test scores by comparing schools with enrolment just below and just above 40 students. More precisely, such a sharp \ac{RD} design estimates the average causal effect for schools with enrolment at the threshold---those who are at the margin of becoming eligible for the small classroom treatment.

But what if compliance with the Maimonides' rule is imperfect? That is, what if some schools opt for small classrooms even if their cohort size falls below the threshold, and others don't open two classrooms even if their cohort size falls above it? In such fuzzy \ac{RD} design, the treatment probability would still increase as we cross the threshold, but it doesn't jump all the way from zero to one as in a sharp design. \citet{htv01} adapted the LATE framework to show that, if in this case we use the class size running variable as an \emph{instrument} for the small classroom treatment (again restricting the analysis to schools with enrolment close to the eligibility threshold), we estimate a LATE: the average treatment effect for compliers---schools at the threshold of eligibility who comply with the treatment assignment rule. The conditions for this result very much mirror the LATE assumptions in \Cref{sec:late}. This influential result shows that \ac{RD} designs and \ac{IV} designs are close cousins, and helped bring about an explosion of \ac{RD} studies in recent years. Methodological work by the laureates also played an important role in boosting the popularity of \ac{RD}. Their contributions range from developing a procedure for selecting the estimation window, formalizing what ``close to the eligibility threshold'' means in practice \citep{ik12} to developing a framework for extrapolating the treatment effects to those away from the cutoff \citep{AnRo15}, and to adapting the LATE framework to a closely related regression kink design, where a continuous treatment variable is a piecewise linear function of a running variable \citep{clpw15}, again with possibly imperfect compliance.

The basic approach of Imbens and Angrist---use the potential outcome framework, keep treatment effect heterogeneity unrestricted, and separate the role of any random variation provided by the natural experiment from other substantive restrictions that are needed---has been fruitfully applied by other researchers in many other contexts besides \ac{RD}. \cite{angrist1998} used the framework to interpret certain fixed-effect regressions. The modern literature on differences-in-differences methods is still exploring the subtle conceptual issues that this approach highlights \citep[see, e.g.][among many others]{dCDH20,SuAb20,contaminationbias}.

Finally, in a series of influential papers,  \cite{heckman/vytlacil:05,heckman/vytlacil_mte1,heckman/vytlacil_mte2} develop a broader framework of \acp{MTE}: the effects for individuals at particular values of an unobserved preference for participation in treatment.
Building on work by \citet{bjorklundmoffitt}, \cite{heckman/vytlacil:05} show how the LATE result fits within this framework. \cite{vytlacil2002} shows that the \ac{MTE} framework is formally equivalent to the LATE model of \citet{ia94}, with a binary or multivalued instrument. In particular, the key monotonicity assumption is equivalent to additive separability between instruments and unobservables in a latent index model of treatment choice.  In the latent index setup, LATE can be seen as an average of \acp{MTE} over a specific range of unobserved preferences, highlighting that effects for compliers for a particular instrument may differ from effects for individuals affected by alternative hypothetical policy changes.

\section{Conclusion}

The 2021 Nobel laureates helped shape modern applied research in labour economics and beyond. A focus on clear research designs exploiting natural experiments led David Card to several empirical conclusions---on immigration effects, firm monopsony power, and educational quality---which challenged conventional wisdom and fueled large bodies of follow-up literatures.
Joshua Angrist and Guido Imbens showed that instrumental variables estimators retain internal validity even when restrictive models for the outcome are relaxed. Their LATE theorem, rooted in a flexible potential outcomes framework, is underpinned by clear and interpretable assumptions and has similarly led to a large body of subsequent applied econometric research.

More broadly, this methodological and empirical focus on clear research designs, and on understanding what core assumptions underlie its internal validity helps portability. Follow-up studies can be conducted in other contexts, probing external validity and replicability. The idea that the research design needs to be tied to the institutional features or other forces driving treatment assignment helps to limit specification searches. The laureates' work also showed how, for more complicated questions, simple research designs can be complemented by careful modeling.

Our brief review focuses on these core contributions of the laureates and necessarily omits many other important contributions. As examples, we have not discussed Card's empirical studies of firm wage-setting \citep[e.g.][]{chk2013,cck2016,cchk2018}; Angrist's methodological work on leveraging the randomness in centralized assignment mechanisms \citep[e.g.][]{mdrd,mdrd2,vam2}; or Imbens' econometric contributions to the estimation of treatment effects under conditional random assignment \citep[e.g.][]{imbens00,hir03,ai06}, or to generalizing the difference-in-differences framework \citep[e.g.][]{c_in_c}. The laureates also made many contributions that address several technical implementation issues that come up in design-based studies, such as how to deal with many weak instruments \citep{AnKr95,aik99}. Not least among these other contributions is the laureates' generosity and dedication when helping their colleagues or advising their students, a trait that we have had the privilege to benefit from first-hand.

\clearpage
\singlespacing
\bibliographystyle{jpe}
\bibliography{SJE}

\end{document}